\documentclass[conference]{IEEEtran}
\usepackage{textcomp}
\usepackage[utf8]{inputenc}
\usepackage{amsmath}
\usepackage{amssymb}
\usepackage{graphicx}

\makeatletter
\IEEEoverridecommandlockouts

\usepackage{cite}
\usepackage{amsfonts}\usepackage{algorithmic}
\usepackage{textcomp}
\usepackage{xcolor}
\def\BibTeX{{\rm B\kern-.05em{\sc i\kern-.025em b}\kern-.08em
    T\kern-.1667em\lower.7ex\hbox{E}\kern-.125emX}}

\makeatother

\begin{document}
\title{Optimizing Antenna Coding for Pixel Antenna Empowered SISO-OFDM Systems\\
}
\author{\IEEEauthorblockN{Tianrui Qiao\textsuperscript{1}, Shanpu Shen\textsuperscript{2,3},
Yijun Chen\textsuperscript{1}, and Ross Murch\textsuperscript{1}} \IEEEauthorblockA{\textsuperscript{1}Department of Electronic and Computer Engineering,
The Hong Kong University of Science and Technology, Hong Kong} \IEEEauthorblockA{\textsuperscript{2}State Key Laboratory of Internet of Things for
Smart City, University of Macau, Macau} \IEEEauthorblockA{\textsuperscript{3}Department of Electrical and Computer Engineering,
University of Macau, Macau} \IEEEauthorblockA{Email: shanpushen@um.edu.mo}}
\maketitle
\begin{abstract}
This work investigates antenna coding optimization to enhance the
channel capacity of single-input single-output orthogonal frequency
division multiplexing (SISO-OFDM) systems empowered by highly reconfigurable
pixel antennas. We first introduce the model for pixel antenna empowered
SISO-OFDM systems using a beamspace channel representation. We next
formulate the problem to maximize the channel capacity through jointly
optimizing antenna coding and the power allocation across subcarriers
and solve it by Successive Exhaustive Boolean Optimization (SEBO)
and water-filling (WF) algorithm. To reduce computational complexity,
a codebook-based approach is also proposed for antenna coding optimization.
Simulation results show that the channel capacity of SISO-OFDM system
across all signal-to-noise-ratio (SNR) regions considered can be enhanced
through leveraging pixel antennas as compared to using conventional
antenna with fixed configuration. This result demonstrates the effectiveness
of antenna coding technology empowered by pixel antenna in enhancing
SISO-OFDM systems.
\end{abstract}


\section{Introduction}

The demanding requirements of the performance of the sixth generation
(6G) wireless communication, such as extremely high data rates, low
latency, and massive connectivity, promote the exploration of new
paradigms in antenna technology \cite{Shen_Antenna_Coding}. Reconfigurable
antennas are one of the promising approaches to breakthrough the performance
bottlenecks of existing wireless communications \cite{KitWC2025}.
Compared to conventional antennas with fixed configuration, reconfigurable
antennas can flexibly adjust the configuration and antenna characteristics
to adapt to the dynamic channel. Therefore, the system performance
can be significantly enhanced to meet various requirements \cite{RA3}.

Pixel antennas are a highly reconfigurable antenna technology and
its concept is based on discretizing a continuous radiation surface
into small sub-wavelength elements called pixels and connecting them
by radio-frequency (RF) switches \cite{Shen_Antenna_Coding}. Through
changing the states of switches between pixels, the electromagnetic
characteristics of pixel antennas can be flexibly reconfigured, including
radiation pattern, operating frequency, and polarization. In the past
decades, various pixel antennas have been designed to achieve distinct
reconfigurability \cite{PixelRA1}, \cite{pixelRA2}, \cite{PixelRA3},
\cite{pixelRA4} and various advanced optimization approaches have
been proposed to accelerate the pixel antennas design and enhance
the performance \cite{SEBO}, \cite{PixelOpti2}, \cite{PixelOpti3},
\cite{PixelOpti4}. However, these previous works are restricted to
the antenna hardware level while the influence of pixel antennas on
wireless communications at the system level has not been investigated
until recent years. A preliminary attempt to utilize pixel antennas
in wireless communication systems is the fluid antenna system (FAS),
where the antenna position can be adjusted to enhance the system performance
\cite{KitFAS}, \cite{KitFAS2}. By optimizing states of switches,
pixel antennas can mimic the position adjustable capability of FAS
\cite{ZhangFAS}, which reveals the possibility of pixel antennas
for improving wireless communications.

To further leverage the significant potential of pixel antenna, a
novel technology, referred to as antenna coding, empowered by pixel
antennas has been proposed, which generalizes the radiation pattern
reconfigurability of pixel antenna \cite{Shen_Antenna_Coding}. The
states of switches are represented by binary variables, which are
referred to as antenna coders. By optimizing the antenna coder, the
radiation pattern of the pixel antenna can be accordingly optimized
to enhance various wireless systems including single-input single-output
(SISO) \cite{Shen_Antenna_Coding}, multiple-input multiple-output
(MIMO) \cite{Shen_Antenna_Coding}, MIMO wireless power transfer (WPT)
\cite{WPT_coding1}, \cite{WPT_coding}, and multi-user (MU) communication
systems \cite{MU_coding}, \cite{MU_coding2}. In addition, spatial
multiplexing gain can be achieved by modulating information onto orthogonal
basis radiation patterns of pixel antennas using antenna coding \cite{Han_coding}.
However, these works \cite{Shen_Antenna_Coding}, \cite{WPT_coding1},
\cite{WPT_coding}, \cite{MU_coding}, \cite{MU_coding2}, \cite{Han_coding}
are limited to single-carrier transmission and overlook exploiting
antenna coding to enhance the multi-carrier transmission.

Therefore, in this work, we investigate antenna coding based on pixel
antennas to enhance multi-carrier transmission with orthogonal frequency
division multiplexing (OFDM). Specifically, we consider a SISO-OFDM
system which utilizes a pixel antenna at the receiver, and optimize
the antenna coder to enhance channel capacity. The contributions of
this work include the following. \textit{First}, we propose a SISO-OFDM
system model with antenna coding empowered by pixel antennas in beamspace.
\textit{Second}, we formulate the SISO-OFDM channel capacity maximization
problem and provide an optimization scheme to jointly optimize the
antenna coder and the power allocated to each subcarrier. \textit{Third},
we design a codebook to facilitate the antenna coding optimization
for the reduction of computational complexity. \textit{Fourth}, we
present simulation results to evaluate the performance of the pixel
antenna empowered SISO-OFDM system and demonstrate the effectiveness
of the pixel antenna.

\textit{Notation}: Bold lower and upper case letters denote vectors
and matrices, respectively. Upper case letters in calligraphy represent
sets. Letters not in bold font are scalars. $\mathbb{R}$ and $\mathbb{C}$
denote real and complex number sets, respectively. $j=\sqrt{-1}$
is the imaginary unit. $\mathcal{CN}\left(\mu,\sigma^{2}\right)$
is a complex Gaussian distribution with mean $\mu$ and standard deviation
$\sigma$. $\{a\}^{+}$ and $\left|a\right|$ are the positive part
and the absolute value of a scalar $\mathit{a}$, respectively. $\mathrm{diag}\left(a_{1},\ldots,a_{N}\right)$
is a diagonal matrix with diagonal entries being $a_{1},\ldots,\mathit{a_{N}}$.
$\lVert\mathbf{a}\rVert$ and $\left[\mathbf{a}\right]_{q}$ are the
$l$-2 norm and the $q$th entry of a vector $\mathbf{a}$, respectively.
$\left[\mathbf{A}\right]_{:,i}$ is the $i$th column of a matrix
$\mathbf{A}$. $(\cdot)^{\mathit{T}}$, $(\cdot)^{\mathit{H}}$, $(\cdot)^{\ast}$
denote the transpose, conjugate transpose, and conjugate, respectively.
$\mathbb{E}\left[\cdot\right]$ represents the expectation. 

\section{Antenna Coding Based on Pixel Antenna}

In this section, we briefly introduce the model for a pixel antenna
and the antenna coding technology.
\begin{figure}[t]
\centering{}\includegraphics[scale=0.41]{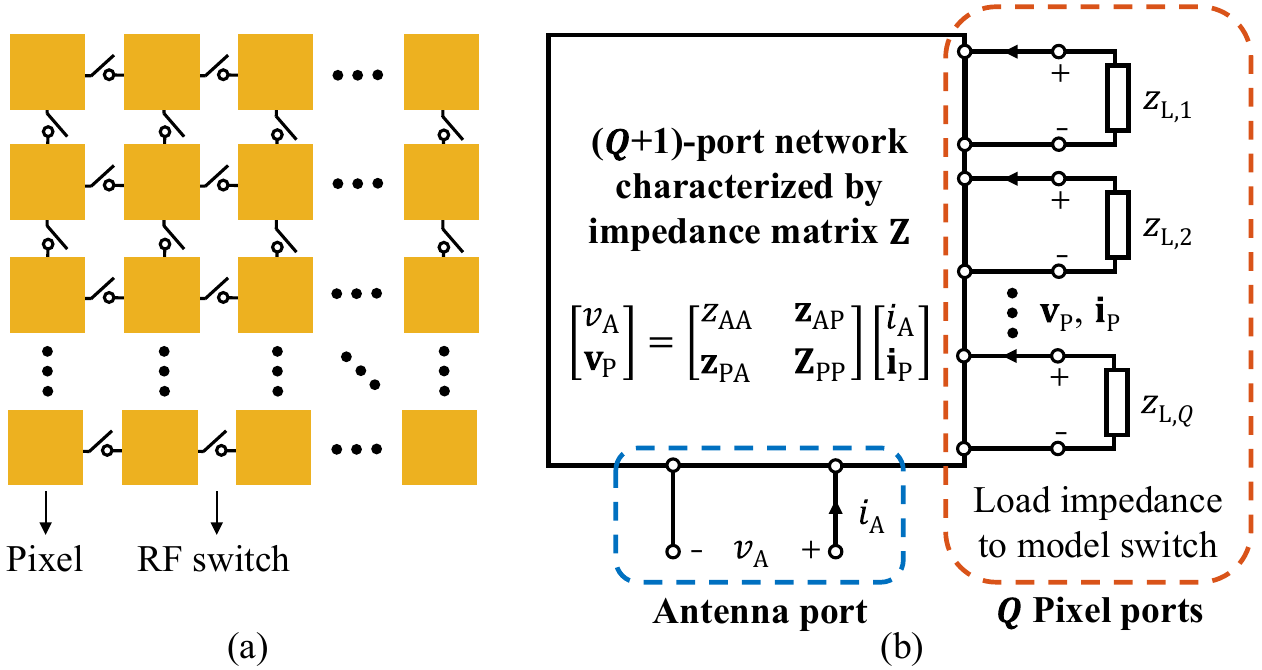} \caption{(a) Schematic diagram and (b) equivalent multiport network model of
pixel antenna.}
\end{figure}

The basic schematic diagram of a pixel antenna is presented in Fig.
1 (a), where a metallic radiation surface is discretized into small
pixels with size much smaller than a wavelength. The adjacent pixels
are connected by $Q$ RF switches and through flexibly tuning their
ON and OFF states, the pixel antenna becomes highly reconfigurable.
We use a binary variable $b_{q}\in\left\{ 0,1\right\} $ to represent
the state of switch connected to the $q$th pixel port, where $0$
and $1$ represent the ON-state with load impedance $z_{{\mathrm{L}},q}=0$
and the OFF-state with load impedance $z_{{\mathrm{L}},q}=\infty$,
respectively. The states of all $Q$ switches $b_{q}$, $\forall q\in\mathcal{Q}=\left\{ 1,\ldots,Q\right\} $,
are grouped into a vector $\mathbf{b}=\left[b_{1},\ldots,b_{Q}\right]^{T}\in\left\{ 0,1\right\} ^{Q\times1}$,
referring to as antenna coder.

According to the multiport network theory, a pixel antenna with a
single feeding port and $Q$ switches can be equivalently modeled
as a $(Q+1)$-port network, consisting of one antenna port connected
to the external source and $Q$ pixel ports, as illustrated in Fig.
1 (b). We characterize the $(Q+1)$-port circuit network by an impedance
matrix $\mathbf{Z}=[z_{\mathrm{AA}},\mathbf{z}_{\mathrm{AP}};\mathbf{z}_{\mathrm{PA}},\mathbf{Z}_{\mathrm{PP}}]\in\mathbb{C}^{\left(Q+1\right)\times\left(Q+1\right)}$,
where $z_{\mathrm{AA}}\in\mathbb{C}$, $\mathbf{z}_{\mathrm{AP}}\in\mathbb{C}^{1\times Q}$,
and $\mathbf{Z}_{\mathrm{PP}}\in\mathbb{C}^{Q\times Q}$ indicate
the self-impedance of the antenna port, the trans-impedance between
the antenna port and pixel ports, and the self-impedance of the pixel
ports, respectively, and $\mathbf{z}_{\mathrm{PA}}=\mathbf{z}_{\mathrm{AP}}^{T}$.
As presented in Fig. 1 (b), the current at antenna port $i_{\mathrm{A}}\in\mathbb{C}$
and the currents at pixel ports ${\mathbf{i}}_{\mathrm{P}}\in\mathbb{C}^{Q\times1}$
can be related to the voltage at antenna port $v_{\mathrm{A}}\in\mathbb{C}$
and the voltages at pixel ports ${\mathbf{v}}_{\mathrm{P}}\in\mathbb{C}^{Q\times1}$
by the impedance matrix $\mathbf{Z}$ as
\begin{equation}
\left[\begin{array}{c}
v_{\mathrm{A}}\\
{\mathbf{v}}_{\mathrm{P}}
\end{array}\right]=\left[\begin{array}{cc}
z_{\mathrm{AA}} & \mathbf{z}_{\mathrm{AP}}\\
\mathbf{z}_{\mathrm{PA}} & \mathbf{Z}_{\mathrm{PP}}
\end{array}\right]\left[\begin{array}{c}
i_{\mathrm{A}}\\
{\mathbf{i}}_{\mathrm{P}}
\end{array}\right].
\end{equation}
Further leveraging the relationship ${\mathbf{v}}_{\mathrm{P}}=-{\mathbf{Z}}_{\mathrm{L}}\left(\mathbf{b}\right){\mathbf{i}}_{\mathrm{P}}$,
we have 
\begin{equation}
\mathbf{i}_{\mathrm{P}}(\mathbf{b})=-\left(\mathbf{Z}_{\mathrm{PP}}+\mathbf{Z}_{\mathrm{L}}\left(\mathbf{b}\right)\right)^{-1}\mathbf{z}_{\mathrm{PA}}i_{\mathrm{A}},\label{eq:current}
\end{equation}
where $\mathbf{Z}_{\mathrm{L}}\left(\mathbf{b}\right)=\mathrm{diag}\left(z_{{\mathrm{L}},1},\ldots,z_{{\mathrm{L}},Q}\right)\in\mathbb{C}^{Q\times Q}$
is a diagonal load impedance matrix. It can be observed from (\ref{eq:current})
that the currents on the pixel ports $\mathbf{i}_{\mathrm{P}}(\mathbf{b})$
are coded by the antenna coder $\mathbf{b}$.

The corresponding coded radiation pattern of the pixel antenna $\mathbf{e}\left(\mathbf{b}\right)\in\mathbb{C}^{2V\times1}$,
containing $\theta$ and $\phi$ polarization components uniformly
sampled over $V$ spatial angles, can be expressed as 
\begin{equation}
\mathbf{e}\left(\mathbf{b}\right)=\mathbf{E}_{\mathrm{oc}}\mathbf{i}\left(\mathbf{b}\right),\label{eq:e(b)}
\end{equation}
where $\mathbf{E}_{\mathrm{oc}}=[\mathbf{e}_{\mathrm{A}}^{\mathrm{oc}},\mathbf{e}_{\mathrm{P},1}^{\mathrm{oc}},\ldots,\mathbf{e}_{\mathrm{P},Q}^{\mathrm{oc}}]\in\mathbb{C}^{2V\times\left(Q+1\right)}$
denotes the open-circuit radiation pattern matrix, with $\mathbf{e}_{\mathrm{A}}^{\mathrm{oc}}\in\mathbb{C}^{2V\times1}$
and $\mathbf{e}_{\mathrm{P},q}^{\mathrm{oc}}\in\mathbb{C}^{2V\times1}$,
$\forall q\in\mathcal{Q}$, representing the radiation patterns (including
the dual polarization components) of the antenna port and the $q$th
pixel port excited by a unit current while all the other ports are
open-circuit, respectively. $\mathbf{e}\left(\mathbf{b}\right)$ is
the superposition of $\mathbf{e}_{\mathrm{A}}^{\mathrm{oc}}$ and
$\mathbf{e}_{\mathrm{P},q}^{\mathrm{oc}}$, $\forall q\in\mathcal{Q}$,
weighted by the currents $\mathbf{i}\left(\mathbf{b}\right)=[i_{\mathrm{A}};\mathbf{i}_{\mathrm{P}}\left(\mathbf{b}\right)]\in\mathbb{C}^{(Q+1)\times1}$.
Thus, through optimizing the antenna coder $\mathbf{b}$ among all
the $2^{Q}$ possible switch configurations, the radiation pattern
of the pixel antenna $\mathbf{e}\left(\mathbf{b}\right)$ can be flexibly
reconfigured to adapt to the wireless propagation channel, potentially
providing more flexibility to the beam control and thus improving
the performance of the wireless communication system.

\section{Pixel Antenna Empowered SISO-OFDM System}

\begin{figure}[t]
\centering{}\includegraphics[scale=0.4]{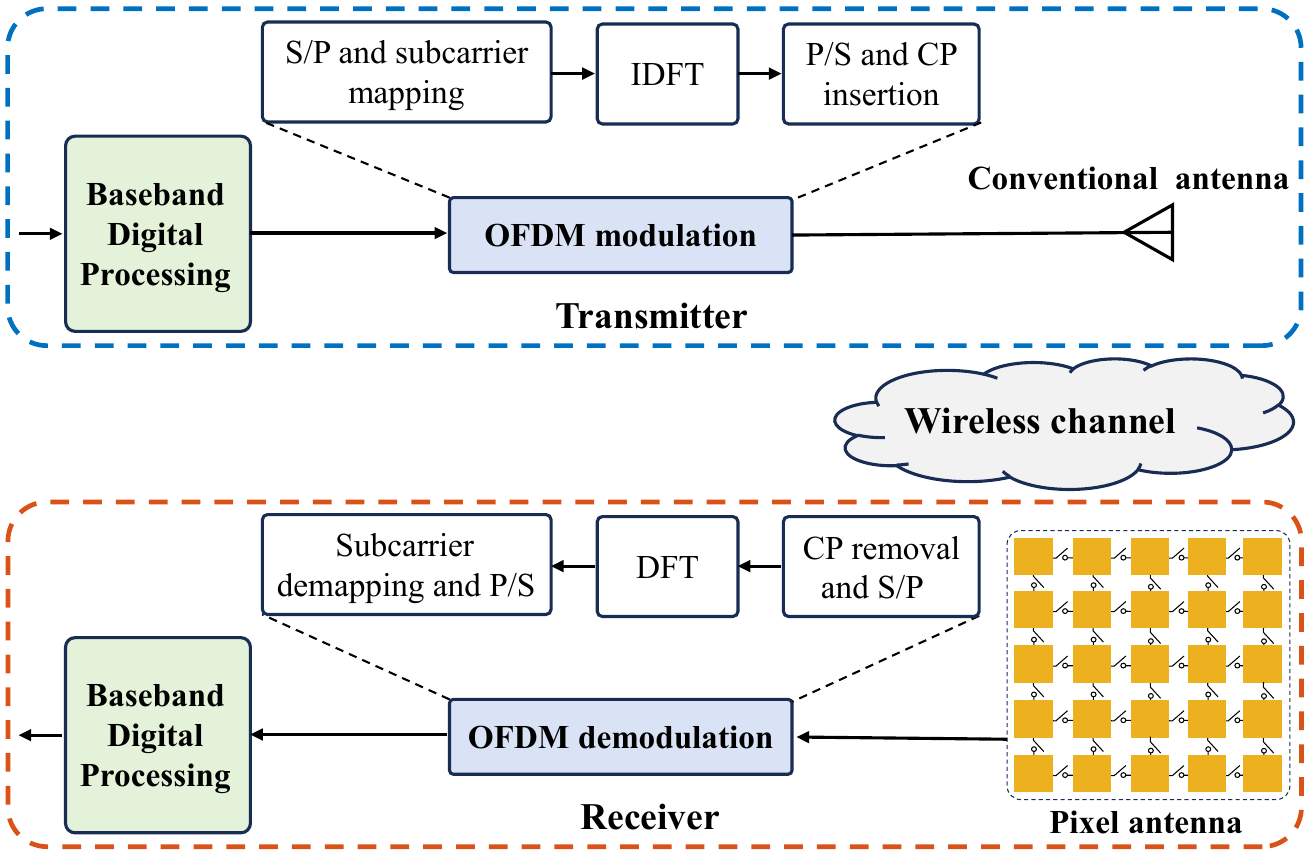} \caption{Diagram of a SISO-OFDM system with an antenna having fixed configuration
at the transmitter and a pixel antenna at the receiver.}\label{fig:SISO-OFDM}
\end{figure}

To reveal the advantages of enhancing wireless communication systems
by using pixel antenna with antenna coding technology, we consider
a SISO-OFDM system with a conventional antenna with fixed configuration
at the transmitter and a pixel antenna at the receiver, as illustrated
in Fig. \ref{fig:SISO-OFDM}. We assume that the SISO-OFDM pixel antenna
system uses $K$ subcarriers for transmission. The $k$th subcarrier
frequency $f_{k}$, $\forall k\in\mathcal{K}=\left\{ 1,\ldots,K\right\} $,
is obtained by $f_{k}=f_{c}+\Delta f\left(k-1-\left(K-1\right)/2\right)$,
where $f_{c}$ is the center frequency and $\Delta f=B/K$ refers
to the subcarrier spacing with $B$ denoting the bandwidth \cite{Hongyu_BDRIS_OFDM}.
We assume that the wireless propagation channel of the SISO-OFDM system
experiences $L$-tap frequency-selective fading while the pixel antenna
is frequency-flat, i.e. the frequency response of the pixel antenna
is constant within the bandwidth $B$. The length of cyclic prefix
(CP) is assumed to be sufficiently long so that there is no inter-symbol
interference. 

Leveraging the beamspace channel representation, the channel for the
pixel antenna empowered SISO-OFDM system at the $k$th subcarrier,
$\forall k\in\mathcal{K}$, can be modeled as 
\begin{equation}
h_{k}\left(\mathbf{b}\right)=\mathbf{e}^{T}\left(\mathbf{b}\right)\mathbf{H}_{\mathrm{V},k}\mathbf{e}_{\mathrm{T}},\label{eq:hk_ori}
\end{equation}
where the radiation pattern of pixel antenna at the receiver $\mathbf{e}(\mathbf{b})$
and the radiation pattern of the conventional antenna at the transmitter
$\mathbf{e}_{T}\in\mathbb{C}^{2V\times1}$ are normalized, satisfying
$\left\Vert \mathbf{e}(\mathbf{b})\right\Vert =1$ and $\left\Vert \mathbf{e}_{T}\right\Vert =1$,
respectively. $\mathbf{H}_{\mathrm{V},k}\in\mathbb{C}^{2V\times2V}$
is the virtual channel at the $k$th subcarrier frequency which contains
the channel gain from each angle of departure to each angle of arrival,
with the angles measured within the local coordinates of the transmitter
and the receiver. The frequency-domain virtual channel $\mathbf{H}_{\mathrm{V},k}$
can be obtained from the time domain through discrete Fourier transform
(DFT) as $\mathbf{H}_{\mathrm{V},k}=\sum_{l=1}^{L}\tilde{\mathbf{H}}_{\mathrm{V},l}\exp\left(-j2\pi\left(k-1\right)\left(l-1\right)/K\right)$,
where $\tilde{\mathbf{H}}_{\mathrm{V},l}\in\mathbb{C}^{2V\times2V}$,
$\forall l\in\mathcal{L}=\left\{ 1,\ldots,L\right\} $, is the time-domain
virtual channel at the $l$th tap and $\{\tilde{\mathbf{H}}_{\mathrm{V},l}\mid\forall l\in\mathcal{L}\}$
are assumed to be independent random matrices \cite{SNR1}. We consider
a rich scattering environment with Rayleigh fading so that each entry
in the time-domain virtual channel $\tilde{\mathbf{H}}_{\mathrm{V},l}$,
$\forall l\in\mathcal{L}$, is independently and identically distributed
(i.i.d.) following complex Gaussian distribution $\mathcal{CN}\left(0,1/L\right)$.
Accordingly, each entry in the frequency-domain virtual channel $\mathbf{H}_{\mathrm{V},k}$
satisfies i.i.d. $\mathcal{CN}\left(0,1\right)$.

To further simply the system model (\ref{eq:hk_ori}) and explicitly
reveal the benefits of using pixel antenna, we decompose the open-circuit
radiation pattern matrix through singular value decomposition as $\mathbf{E}_{\mathrm{oc}}=\mathbf{USV}^{H}$,
where $\mathbf{U}\in\mathbb{C}^{2V\times r}$ and $\mathbf{V}\in\mathbb{C}^{(Q+1)\times r}$
are semi-unitary matrices with $r$ denoting the effective rank of
$\mathbf{E}_{\mathrm{oc}}$ and $\mathbf{S}\in\mathbb{R}^{r\times r}$
is a diagonal matrix collecting the $r$ nonzero singular values.
$r$ can be regarded as the number of orthonormal basis radiation
patterns that can be excited by the pixel antenna, which can also
be referred to as the effective aerial degrees-of-freedom (EADoF)
\cite{Shen_Antenna_Coding}. On top of the decomposition, the coded
radiation pattern of the pixel antenna (\ref{eq:e(b)}) can be rewritten
as 
\begin{equation}
\mathbf{e}\left(\mathbf{b}\right)=\mathbf{USV}^{H}\mathbf{i}\left(\mathbf{b}\right)=\mathbf{U}\mathbf{w}^{\ast}(\mathbf{b}),\label{eq:pattern_coder}
\end{equation}
where we define $\mathbf{w}(\mathbf{b})\triangleq\mathbf{SV}^{T}\mathbf{i}^{\ast}\left(\mathbf{b}\right)\in\mathbb{C}^{r\times1}$
as the pattern coder satisfying $\left\Vert \mathbf{w}(\mathbf{b})\right\Vert =1$,
which indicates how the $r$ orthonormal basis radiation patterns,
i.e. each of the $r$ columns in $\mathbf{U}$, are linearly coded
to synthesize $\mathbf{e}\left(\mathbf{b}\right)$. Utilizing the
pattern coder and substituting (\ref{eq:pattern_coder}) into (\ref{eq:hk_ori}),
the beamspace channel model for the pixel antenna empowered SISO-OFDM
system at the $k$th subcarrier $h_{k}\left(\mathbf{b}\right)$ can
be rewritten as
\begin{equation}
h_{k}\left(\mathbf{b}\right)=\mathbf{w}^{H}(\mathbf{b})\bar{\mathbf{h}}_{k},\forall k\in\mathcal{K},\label{eq:hk}
\end{equation}
where $\bar{\mathbf{h}}_{k}\triangleq\mathbf{U}^{T}\mathbf{H}_{\mathrm{V},k}\mathbf{e}_{\mathrm{T}}\in\mathbb{C}^{r\times1}$
denotes the equivalent channel matrix in beamspace at the $k$th subcarrier
with each entry following i.i.d. $\mathcal{CN}\left(0,1\right)$ because
of the orthonormality among the basis radiation patterns in $\mathbf{U}$
and $\left\Vert \mathbf{e}_{T}\right\Vert =1$. It can be revealed
from the definition of $\bar{\mathbf{h}}_{k}$ that the pixel antenna
functions as a $r$-antenna array, providing additional spatial degrees-of-freedom.
In addition, due to the compact size of $\bar{\mathbf{h}}_{k}$ as
generally $r\ll V$, the required computational complexity for designing
the codebook for antenna coding optimization can be effectively reduced.

Based on (\ref{eq:hk}), the received signal at the $k$th subcarrier
$y_{k}\in\mathbb{C}$, $\forall k\in\mathcal{K}$, can be expressed
as 
\begin{equation}
y_{k}=h_{k}\left(\mathbf{b}\right)x_{k}+n_{k}=\mathbf{w}^{H}(\mathbf{b})\bar{\mathbf{h}}_{k}x_{k}+n_{k},\label{eq:yk}
\end{equation}
where $x_{k}\in\mathbb{C}$ is the transmitted signal at the $k$th
subcarrier and $n_{k}\in\mathbb{C}$ following $\mathcal{CN}\left(0,N_{0}\right)$
indicates the spectrally-uncorrelated additive white Gaussian noise
at the $k$th subcarrier with noise power $N_{0}$. The system model
can be considered as the frequency response of the time-domain beamspace
channel as introduced in \cite{Han_OFDM}.

\section{Antenna Coding Optimization}

In this section, we first formulate and solve the antenna coding optimization
problem to maximize the channel capacity realized by the pixel antenna
empowered SISO-OFDM system and then propose a codebook-based approach
to optimize the antenna coder for reducing the computational complexity.

\subsection{Formulating and Solving Optimization}

We assume the channel state information (CSI) $\bar{\mathbf{h}}_{k}$,
$\forall k\in\mathcal{K}$, is perfectly known. Leveraging (\ref{eq:hk})
and (\ref{eq:yk}), the optimization problem to maximize the channel
capacity averaged over all the $K$ subcarriers of the pixel antenna
empowered SISO-OFDM system can be formulated as 
\begin{align}
\underset{\{P_{k}\}_{\forall k\in\mathcal{K}},\mathbf{b}}{\mathrm{max}}\quad & \frac{1}{K}\underset{k\in\mathcal{K}}{\sum}\log_{2}\left(1+\frac{P_{k}}{N_{0}}\left|\mathbf{w}^{H}(\mathbf{b})\bar{\mathbf{h}}_{k}\right|^{2}\right)\label{eq:ori_opti_obj}\\
\mathrm{s.t.}\quad\quad\,\, & \underset{k\in\mathcal{K}}{\sum}P_{k}\leq P,\label{eq:ori_opti_cons1}\\
 & \left[\mathbf{b}\right]_{q}\in\left\{ 0,1\right\} ,\forall q\in\mathcal{Q},\label{eq:ori_opti_cons2}
\end{align}
where $P$ represents the total transmit power and $P_{k}$, $\forall k\in\mathcal{K}$,
denotes the transmit power at the $k$th subcarrier. 

The problem (\ref{eq:ori_opti_obj}) to (\ref{eq:ori_opti_cons2})
jointly optimizes the antenna coder and the power allocation across
all the $K$ subcarriers. To solve it, for given channel $\{h_{k}(\mathbf{b})=\mathbf{w}^{H}(\mathbf{b})\bar{\mathbf{h}}_{k}\mid\forall k\in\mathcal{K}\}$
with arbitrary antenna coder $\mathbf{b}$, the optimal power allocation
at each subcarrier that maximizes the channel capacity can be obtained
by the water-filling (WF) algorithm, expressed as 
\begin{equation}
P_{k}^{\star}(\mathbf{b})=\left\{ \mu-\frac{N_{0}}{\left|\mathbf{w}^{H}(\mathbf{b})\bar{\mathbf{h}}_{k}\right|^{2}}\right\} ^{+},\forall k\in\mathcal{K},
\end{equation}
where $\mu$ is selected to satisfy the total transmitted power constraint
$\sum_{k\in\mathcal{K}}P_{k}^{\star}(\mathbf{b})=P$. With the WF
power allocation as a function of $\mathbf{b}$, the antenna coding
design problem (\ref{eq:ori_opti_obj}) to (\ref{eq:ori_opti_cons2})
can thus be reformulated as 
\begin{align}
\underset{\mathbf{b}}{\mathrm{max}}\quad & \frac{1}{K}\underset{k\in\mathcal{K}}{\sum}\log_{2}\left(1+\frac{P_{k}^{\star}(\mathbf{b})}{N_{0}}\left|\mathbf{w}^{H}(\mathbf{b})\bar{\mathbf{h}}_{k}\right|^{2}\right)\\
\mathrm{s.t.}\quad & \left[\mathbf{b}\right]_{q}\in\left\{ 0,1\right\} ,\forall q\in\mathcal{Q},
\end{align}
which is an NP-hard binary optimization problem. To solve this binary
optimization problem, we apply the Successive Exhaustive Boolean Optimization
(SEBO) algorithm. The main idea of SEBO is to exhaustively search
each block and randomly flip the bits of the antenna coder iteratively
until convergence. More details of SEBO can be found in \cite{SEBO}.

\subsection{Codebook-Based Antenna Coding Optimization}

Considering that the iterative successive exhaustive search required
in the SEBO causes high computational complexity, we design a codebook
to reduce the computational complexity for antenna coding optimization.
Specifically, we define the codebook for antenna coder as $\mathcal{B}\triangleq\{\mathbf{b}_{m}\in\left\{ 0,1\right\} ^{Q\times1}\mid\forall m\in\mathcal{M}=\{1,\ldots,M\}\}$
with $M$ denoting the codebook size. Given the codebook and channel
$\{h_{k}(\mathbf{b})\mid\forall k\in\mathcal{K}\}$, the optimal antenna
coder $\mathbf{b}^{\star}$ that maximizes the capacity can be found
by searching the codebook, i.e. 
\begin{equation}
\mathbf{b}^{\star}=\underset{\mathbf{b}\in\mathcal{B}}{\mathrm{argmax}}\:\frac{1}{K}\underset{k\in\mathcal{K}}{\sum}\log_{2}\left(1+\frac{P_{k}^{\star}(\mathbf{b})}{N_{0}}\left|\mathbf{w}^{H}(\mathbf{b})\bar{\mathbf{h}}_{k}\right|^{2}\right).\label{eq:b_star}
\end{equation}

The key issue is to design an appropriate codebook $\mathcal{B}$
such that the ergodic channel capacity of the pixel antenna empowered
SISO-OFDM system can be maximized, which can be equivalently formulated
as
\begin{align}
\underset{\mathbf{b}_{m},\forall m}{\mathrm{max}}\quad & \mathbb{E}\left[\underset{k\in\mathcal{K}}{\sum}\log_{2}\left(1+\frac{P_{k}^{\star}(\mathbf{b}^{\star})}{N_{0}}\left|\mathbf{w}^{H}(\mathbf{b}^{\star})\bar{\mathbf{h}}_{k}\right|^{2}\right)\right]\label{eq:cdb_design1-1}\\
\mathrm{s.t.}\quad\:\, & \left[\mathbf{b}_{m}\right]_{q}\in\left\{ 0,1\right\} ,\forall q\in\mathcal{Q}.\label{eq:cdb_design2-1}
\end{align}
To solve this problem, a training set is constructed as 
\begin{align}
\tilde{\mathcal{H}} & =\left\{ \mathcal{\tilde{H}}^{d}\mid\forall d\in\mathcal{D}=\{1,\ldots,D\}\right\} ,\label{eq:H_time_set}
\end{align}
where $D$ is the number of channel realizations and $\tilde{\mathcal{H}}^{d}$
is the $d$th channel realization containing channels at $L$ delay
taps, written as 
\begin{align}
\tilde{\mathcal{H}}^{d} & =\left\{ \tilde{\mathbf{H}}_{\mathrm{V},1}^{d}\in\mathbb{C}^{2V\times2V},\ldots,\tilde{\mathbf{H}}_{\mathrm{V},L}^{d}\in\mathbb{C}^{2V\times2V}\right\} ,\label{eq:Hs_time-1}
\end{align}
which are assumed following i.i.d. complex Gaussian. Given the training
channel realization in time domain (\ref{eq:Hs_time-1}), we can transfer
it to frequency domain by DFT, so that we have 
\begin{align}
\mathcal{H}^{d} & =\left\{ \mathbf{H}_{\mathrm{V},1}^{d}\in\mathbb{C}^{2V\times2V},\ldots,\mathbf{H}_{\mathrm{V},K}^{d}\in\mathbb{C}^{2V\times2V}\right\} .
\end{align}
Furthermore, given $\mathcal{H}^{d}$, $\forall d\in\mathcal{D}$,
we can make use of \eqref{eq:hk} to find the training set which contains
the corresponding equivalent channel matrix in beamspace, expressed
as 
\begin{equation}
\bar{\mathcal{H}}=\left\{ \bar{\mathbf{H}}^{d}=\left[\bar{\mathbf{h}}_{1}^{d},\ldots,\bar{\mathbf{h}}_{K}^{d}\right]\in\mathbb{C}^{r\times K}\mid\forall d\in\mathcal{D}\right\} ,\label{eq:H_freq_set}
\end{equation}
where we have $\bar{\mathbf{h}}_{k}^{d}\triangleq\mathbf{U}^{T}\mathbf{H}_{\mathrm{V},k}^{d}\mathbf{e}_{\mathrm{T}}$,
$\forall k\in\mathcal{K}$. The training set $\bar{\mathcal{H}}$
can be further partitioned into $M$ subsets $\bar{\mathcal{H}}_{1},\ldots,\bar{\mathcal{H}}_{M}$,
each of which is associated with an antenna coder in the codebook
$\mathcal{B}$. Specifically, $\bar{\mathcal{H}}_{m}$ represents
the nearest neighbor of the $m$th antenna coder $\mathbf{b}_{m}$,
$\forall m\in\mathcal{M}$, and is given by
\begin{align}
\bar{\mathcal{H}}_{m}=\Bigg\{\bar{\mathbf{H}}^{d}\mid\underset{k\in\mathcal{K}}{\sum}C_{k}(\mathbf{b}_{m},\bar{\mathbf{H}}^{d})\geq\nonumber \\
\underset{k\in\mathcal{K}}{\sum}C_{k}(\mathbf{b}_{m^{\prime}},\bar{\mathbf{H}}^{d}),\forall m\neq & m^{\prime},\forall d\in\mathcal{D}\Bigg\},
\end{align}
where $C_{k}(\mathbf{b}_{m},\bar{\mathbf{H}}^{d})$ denotes the capacity
at the $k$th subcarrier, given by
\begin{equation}
C_{k}(\mathbf{b}_{m},\bar{\mathbf{H}}^{d})=\log_{2}\left(1+\frac{P_{k}^{\star}(\mathbf{b}_{m})}{N_{0}}\left|\mathbf{w}^{H}(\mathbf{b}_{m})\left[\bar{\mathbf{H}}^{d}\right]_{:,k}\right|^{2}\right).\label{eq:Ck}
\end{equation}
That is to say, all the channel realizations assigned to the $m$th
partition $\bar{\mathcal{H}}_{m}$ have the highest average channel
capacity across all the $K$ subcarriers with the antenna coder $\mathbf{b}_{m}$
than with any other antenna coders in the codebook. The codebook design
problem (\ref{eq:cdb_design1-1}) to (\ref{eq:cdb_design2-1}) can
thus be equivalently reformulated as
\begin{align}
\underset{\mathbf{b}_{m},\forall m}{\mathrm{max}}\quad & \underset{m\in\mathcal{M}}{\sum}\underset{\bar{\mathbf{H}}^{d}\in\bar{\mathcal{H}}_{m}}{\sum}\underset{k\in\mathcal{K}}{\sum}C_{k}(\mathbf{b}_{m},\bar{\mathbf{H}}^{d})\label{eq:cdb_design1}\\
\mathrm{s.t.}\quad\:\, & \left[\mathbf{b}_{m}\right]_{q}\in\left\{ 0,1\right\} ,\forall q\in\mathcal{Q},\label{eq:cdb_design2}
\end{align}
where the partition of training set is coupled with the optimization
of the antenna coder. Therefore, we apply the generalized Lloyd algorithm
\cite{Lloyd} to solve the problem (\ref{eq:cdb_design1}) to (\ref{eq:cdb_design2})
as briefly summarized below.

\subsubsection{Partition Optimization}

At the $i$th iteration, given the codebook optimized at the $(i-1)$th
iteration $\mathcal{B}^{(i-1)}=\{{\mathbf{b}}_{1}^{(i-1)},\ldots,{\mathbf{b}}_{M}^{(i-1)}\}$,
the $m$th partition of the training set $\bar{\mathcal{H}}_{m}^{(i)}$,
$\forall m\in\mathcal{M}$, is optimized by 
\begin{align}
\bar{\mathcal{H}}_{m}^{(i)}=\Bigg\{\bar{\mathbf{H}}^{d}\mid\underset{k\in\mathcal{K}}{\sum}C_{k}(\mathbf{b}_{m}^{(i-1)},\bar{\mathbf{H}}^{d})\geq\nonumber \\
\underset{k\in\mathcal{K}}{\sum}C_{k}(\mathbf{b}_{m^{\prime}}^{(i-1)},\bar{\mathbf{H}}^{d}),\forall m\neq & m^{\prime},\forall d\in\mathcal{D}\Bigg\}.
\end{align}

\subsubsection{Centroid Optimization}

Given the optimized training set partition $\bar{\mathcal{H}}_{m}^{(i)}$,
the $m$th antenna coder $\mathbf{b}_{m}$, $\forall m\in\mathcal{M}$,
is optimized by
\begin{equation}
\mathbf{b}_{m}^{(i)}=\underset{\left[\mathbf{b}_{m}\right]_{q}\in\left\{ 0,1\right\} ,\forall q}{\mathrm{argmax}}\underset{\bar{\mathbf{H}}^{d}\in\bar{\mathcal{H}}_{m}^{(i)}}{\sum}\underset{k\in\mathcal{K}}{\sum}C_{k}(\mathbf{b}_{m},\bar{\mathbf{H}}^{d}),\label{eq:bm_center}
\end{equation}
which can be solved by SEBO algorithm.

With a random initialization, the above procedures are iteratively
optimized until the antenna coders $\mathbf{b}_{m}$, $\forall m\in\mathcal{M}$,
are converged. At the end of the generalized Lloyd algorithm, the
optimal antenna coders ${\mathbf{b}}_{m}^{\star}$, $\forall m\in\mathcal{M}$,
can be obtained, forming the optimal codebook $\mathcal{B}^{\star}=\{{\mathbf{b}}_{1}^{\star},\ldots,{\mathbf{b}}_{M}^{\star}\}$.
It should be noted that the objective of problem (\ref{eq:cdb_design1})
to (\ref{eq:cdb_design2}) depends on the signal-to-noise-ratio (SNR)
of the SISO-OFDM system\footnote{Here we use the definition of SNR for OFDM \cite{SNR1}, that is the
ratio between the total transmit power and the total noise power across
all subcarriers.}, so that we can design codebooks for different SNR regions. Once
the optimal codebook $\mathcal{B}^{\star}$ is obtained, for a given
channel, the optimal antenna coder $\mathbf{b}^{\star}$ that maximizes
the capacity of the pixel antenna empowered SISO-OFDM system can found
by searching from the optimal codebook $\mathcal{B}^{\star}$. 

\section{Performance Evaluation}

In this section, we evaluate the performance of the proposed pixel
antenna empowered SISO-OFDM system with antenna coding optimization.

\subsection{Simulation Setup}

We assume a 2-D uniform power angular spectrum (PAS) with equally
likely polarization. The angular resolution is set as $5^{\circ}$
so that there are $V=72$ sampled angles. Similar to \cite{Shen_Antenna_Coding},
we consider a pixel antenna at the receiver with a physical aperture
size $0.5\lambda\times0.5\lambda$ operating at 2.4 GHz has one antenna
port and $Q=39$ switches. Its impedance matrix $\mathbf{Z}$ and
open-circuit radiation pattern matrix $\mathbf{E}_{\mathrm{oc}}$
can be obtained by CST studio suite through one-time full-wave electromagnetic
simulation. The EADoF of the pixel antenna $r$, i.e. the rank of
$\mathbf{E}_{\mathrm{oc}}$, is estimated as the number of singular
values of $\mathbf{E}_{\mathrm{oc}}$ that cumulatively contribute
to over 99.8\% of the total radiated power, leading to $r=7$. Regarding
the OFDM settings, we assume that the number of delay taps is $L=4$,
the subcarrier spacing is $\Delta f=312.5$ kHz, the number of subcarriers
is $K=64$, and thus the bandwidth is $B=20$ MHz \cite{Hongyu_BDRIS_OFDM},
\cite{Han_OFDM}. When optimizing the antenna coder by the SEBO algorithm,
the required parameter of block size is set to 10 \cite{Shen_Antenna_Coding}.
When designing the codebook, the size of the training set is determined
as $D=30000$.

\subsection{System Performance}

We evaluate the channel capacity of the pixel antenna empowered SISO-OFDM
system versus SNR, in comparison with the channel capacity of the
conventional SISO-OFDM system, where conventional antennas with fixed
configuration are used at both the transmitter and the receiver. We
design two codebooks, by solving the problem (\ref{eq:cdb_design1})
to (\ref{eq:cdb_design2}), one for SNR of 0 dB and the other for
SNR of 30 dB. When the SNR of the system is higher than 15 dB, the
codebook at high-SNR regime (designed at 30 dB) is leveraged for antenna
coding optimization. Otherwise, the codebook at low-SNR regime (designed
at 0 dB) is used. 

\begin{figure}[t]
\centering{}\includegraphics[width=8cm]{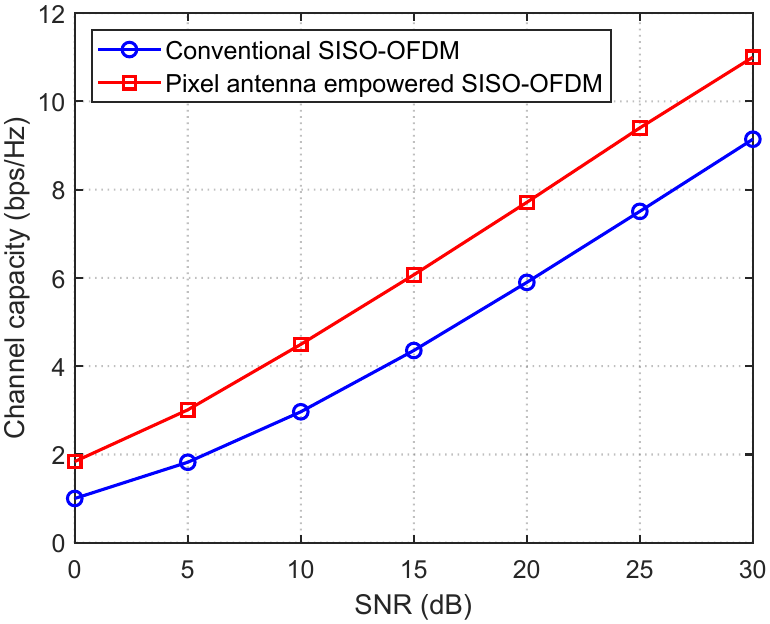} \caption{Channel capacity versus SNR for the proposed pixel antenna empowered
SISO-OFDM system optimized by SEBO algorithm.}\label{fig:CvsSNR}
\end{figure}

In Fig. \ref{fig:CvsSNR}, we show the channel capacity of the pixel
antenna empowered SISO-OFDM system optimized by SEBO algorithm, benchmarked
with the conventional SISO-OFDM system. We can observe from Fig. \ref{fig:CvsSNR}
that using the pixel antenna at the receiver can effectively enhance
the channel capacity of the SISO-OFDM system across all SNRs, compared
with using conventional antenna with fixed configuration. For example,
the channel capacity can be increased by 83\% at low SNR of 0 dB and
by 20\% at high SNR of 30 dB through leveraging the pixel antenna
in comparison with the conventional SISO-OFDM system.

\begin{figure}
\centering{}\includegraphics[width=8cm]{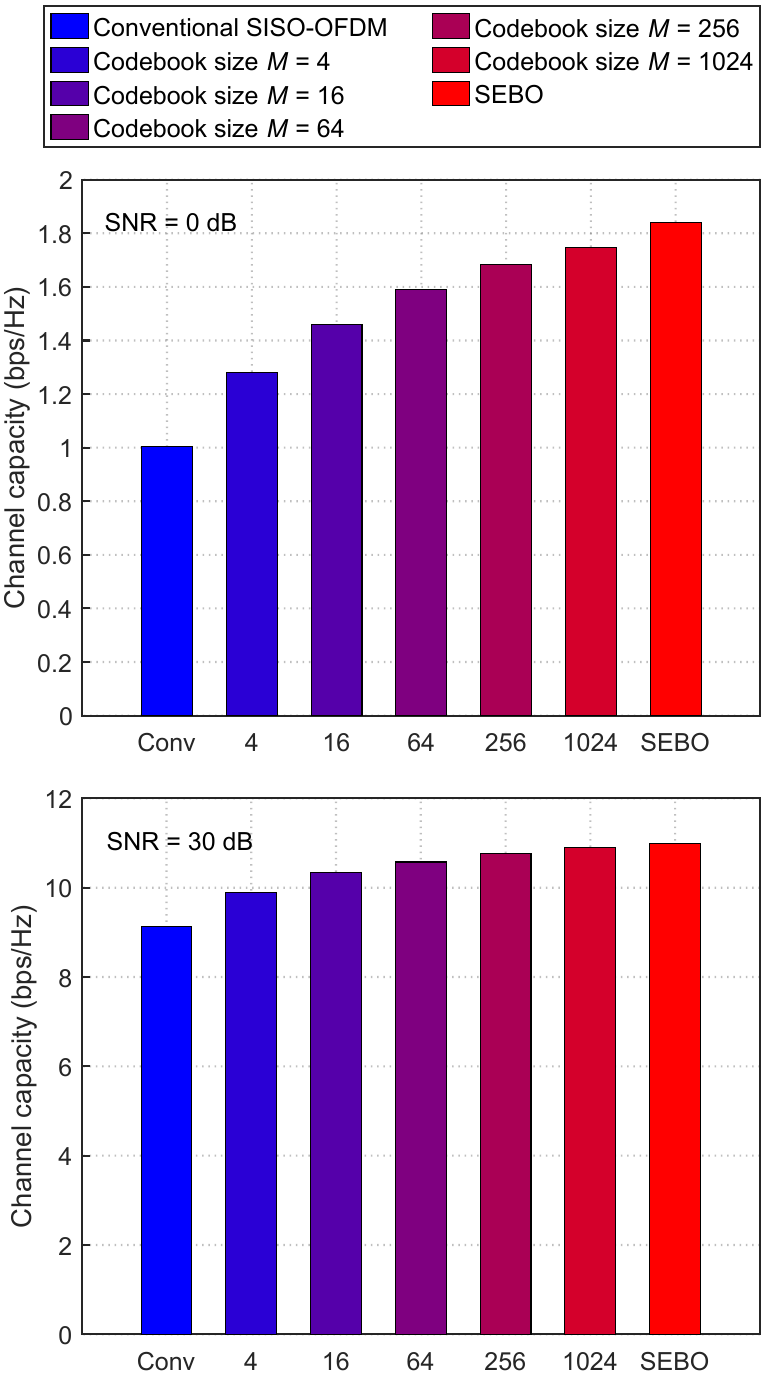} \caption{Channel capacity versus the codebook size for the proposed pixel
antenna empowered SISO-OFDM system at SNR = 0 dB and SNR = 30 dB.}\label{fig:CvsCDBsize}
\end{figure}

In Fig. \ref{fig:CvsCDBsize}, we show the channel capacity of the
pixel antenna empowered SISO-OFDM system versus the codebook size
at low SNR of 0 dB and high SNR of 30 dB. We can find that, with the
enlargement of the codebook size, the capacity of the pixel antenna
empowered SISO-OFDM system optimized by codebook increases and approaches
to that optimized by SEBO. For example, when the codebook has a size
of $M=1024$, the capacity of the system empowered by the pixel antenna
can realize 95\% and 99\% of the performance optimized by SEBO when
SNR = 0 dB and SNR = 30 dB, respectively, while the computational
complexity is significantly reduced. Even when the codebook has a
very small size of $M=4$, the capacity can still be increased by
28\% at low SNR of 0 dB and by 8\% at high SNR of 30 dB. 

\section{Conclusion}

In this work, we investigate antenna coding optimization for pixel
antenna empowered SISO-OFDM systems. The antenna coder and the power
allocation across subcarriers are jointly optimized to maximize the
channel capacity of the pixel antenna empowered SISO-OFDM system by
using the WF power allocation with SEBO algorithm. In addition, to
reduce the computational complexity, a codebook-based approach is
proposed for the antenna coding optimization. Simulation results reveal
that leveraging the pixel antenna at the receiver of the SISO-OFDM
system can effectively improve channel capacity, by up to 83\%. This
is due to its ability to flexibly adjust its radiation pattern to
adapt to the channel, demonstrating the advantages of using pixel
antenna with antenna coding in SISO-OFDM systems. 

\section*{Acknowledgment}

The authors would like to acknowledge the support by the Hong Kong
Research Grants Council for the General Research Fund (GRF) grant
16208124, the support by the Science and Technology Development Fund,
Macau SAR (File/Project no. 001/2024/SKL), the support by University
of Macau (File no. SRG2025-00060-IOTSC), and the support by University
of Macau Development Foundation (UMDF) (File no. UMDF-TISF-I/2026/025/IOTSC).

\end{document}